\documentclass{article}

\PassOptionsToPackage{numbers,compress}{natbib}

\usepackage[main, final]{neurips_2025}

\usepackage[utf8]{inputenc} 
\usepackage[T1]{fontenc}    
\usepackage{hyperref}       
\usepackage{graphicx}
\usepackage{url}            
\usepackage{booktabs}       
\usepackage{amsfonts}       
\usepackage{nicefrac}       
\usepackage{microtype}      
\usepackage{subfig}
\usepackage{float}
\usepackage{xcolor}         

\title{Do Vision-Language Models See Dwarf Galaxies the Way We Do?}

\author{%
  Dimitrios Tanoglidis \\
  Independent Researcher\\
  \texttt{tanoglidisdimitrios@gmail.com} \\
   \And
   Chin Yi Tan \\
  Kavli Institute for Cosmological Physics, University of Chicago \\
  Department of Physics, University of Chicago\\
  NSF-Simons AI Institute for the Sky (SkAI)\\
   \texttt{chinyi@uchicago.edu} \\
   \AND
  Kate Overdeck \\
  Department of Astronomy \& Astrophysics, University of Chicago\\
   NSF-Simons AI Institute for the Sky (SkAI) \\
  \texttt{koverdeck@uchicago.edu} \\
   \And
  Alex Drlica-Wagner \\
  Fermi National Accelerator Laboratory\\
  Kavli Institute for Cosmological Physics, University of Chicago \\
  Department of Astronomy \& Astrophysics, University of Chicago\\
   NSF-Simons AI Institute for the Sky (SkAI)  
   \\
   \texttt{kadrlica@fnal.gov} \\
}

\begin{document}

\maketitle

\begin{abstract}
With the advent of powerful, general-purpose vision-language models (VLMs), there has been growing interest in their potential to assist astronomical discovery, a field characterized by large volumes of image data. In this work, we evaluate VLMs on the challenging task of identifying ultra-faint dwarf galaxy candidates using multi-panel diagnostic images from survey data. We compare model predictions to human annotations from a large-scale citizen science campaign. We find that zero-shot VLMs closely reproduce aggregate human calibration and perform well on less ambiguous cases. However, there is significant variability at the level of individual examples, and attempts to obtain uncertainty estimates (via self-reported confidence or repeated inference) fail to yield reliable and practically useful measures. Our results highlight both the promise and the current limitations of deploying VLMs for large-scale scientific discovery in realistic settings.
\end{abstract}

\section{Introduction}
\label{sec:Introduction}

Large, multimodal, Vision-Language Models (VLMs; e.g.,\ \cite{VLMs_intro}) have demonstrated strong performance across a wide range of scientific benchmarks, multimodal reasoning tasks, and real-world applications(e.g.,\ \cite{Cooper_2024,Kawaharazuka_2024,Li_2025, Yang_2023, MMMU_benchmark}). As general-purpose foundation models\cite{Foundation_Models} pre-trained on large-scale datasets, they can be rapidly adapted to a variety of downstream tasks (e.g., image classification) with minimal task-specific engineering. Importantly, for VLMs, this adaptation can be achieved through natural language instructions, optionally augmented with a small number of example images.

Given their generality and flexibility, there has been growing interest in applying VLMs to astronomical data analysis, where observations are predominantly image-based. Modern astronomical surveys produce vast quantities of imaging data (e.g.,\ \cite{Ivezic_LSST}), and identifying scientifically interesting objects typically requires inspecting, and rejecting, a large number of contaminants (e.g., artifacts or spurious detections).

While custom deep learning models are widely used, they require substantial development effort, large labeled datasets that are often unavailable, and can inherit biases from their training data (especially when synthetic data are used). Citizen science campaigns (e.g., Zooniverse) provide an alternative, but designing, deploying, and analyzing such campaigns also requires substantial effort.

The application of VLMs to astronomy has been explored in recent work \citep{AstroAlertBench_2026, VLM_Ensembles_2026,Drozdova_2025, Koblischke_2025, Sharma_Paperclip,Systematic_VLM_2026,Riggi_2025, Stoppa_2025, Tanoglidis_Jain, Zaman_astrollava}. However, these studies have largely focused on relatively clean, single-image datasets, where the classification task is comparatively straightforward. This leaves open the question of how VLMs perform in more complex, realistic settings that require integrating multiple diagnostic plots.

In this work, we leverage the outputs of a recent citizen science campaign aimed at discovering ultra-faint dwarf galaxies in wide-field survey data. Each candidate is evaluated using multiple diagnostic plots, requiring the integration of complementary visual cues. 
We ask how well general-purpose VLMs match the behavior of citizen scientists, particularly in ambiguous or challenging cases.

We find that VLMs, without task-specific training, recover aggregate human calibration well, but exhibit substantial per-example variability and systematic mismatches in confidence. These results highlight both the promise and the limitations of using general-purpose VLMs for scientific discovery tasks involving complex, multi-view reasoning.

\section{Methods}
\label{se:Methods}

\paragraph{Data:}
We use images and annotations from a recently completed Zooniverse\footnote{\url{https://www.zooniverse.org/}} campaign aimed at discovering ultra-faint Milky Way satellite galaxies in the wide-field DELVE survey, which covers $\sim 20{,}000,\mathrm{deg}^2$ of the southern sky \citep{ADW_Delve_WP}. The observed abundance of dwarf galaxies orbiting the Milky Way provides strong constraints on dark matter models, making their discovery both scientifically important and observationally challenging \cite{Bullock_2017, Nadler_2021, CYTan_2026}. In practice, the search is dominated by large numbers of spurious candidates, necessitating either custom deep learning pipelines or large-scale citizen science efforts to reduce the candidate pool.

\begin{figure}[!ht]
\centering
\includegraphics[width=0.90\textwidth]{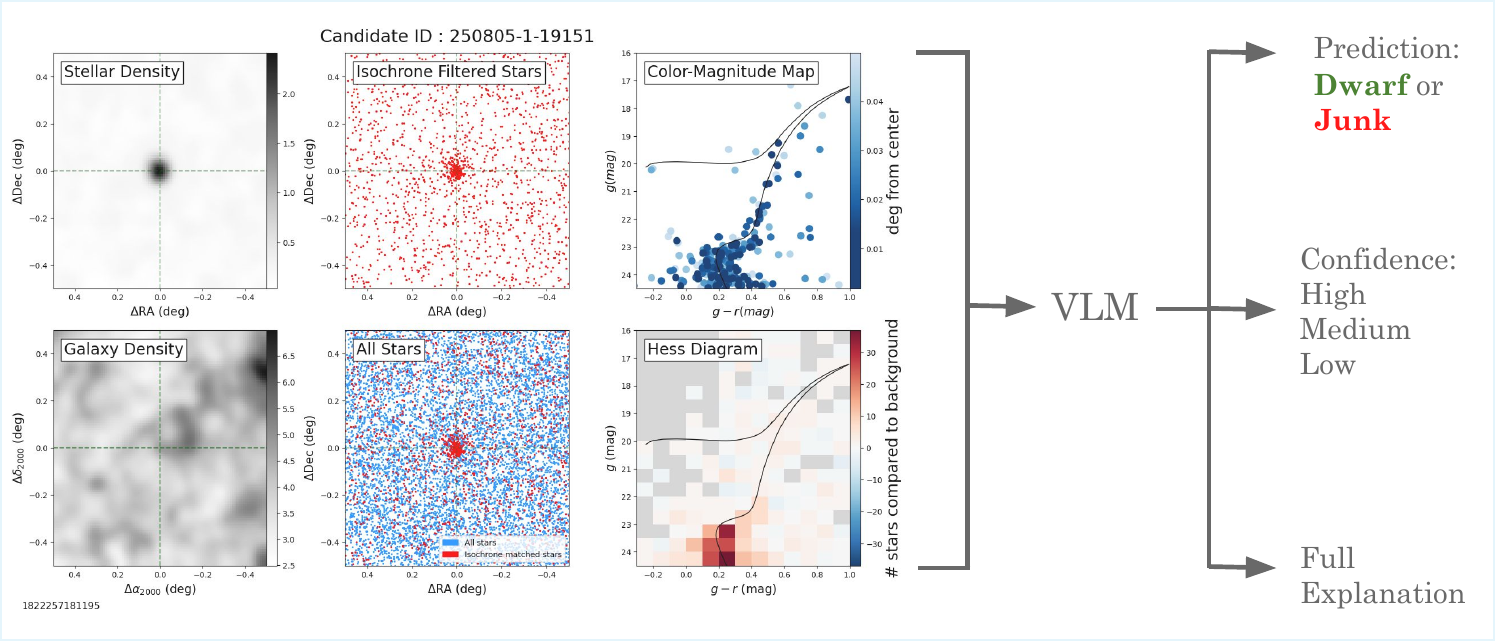}
\caption{Example of the diagnostic panel shown to both Zooniverse volunteers and the VLM. This instance corresponds to the known dwarf galaxy Horologium I, for which all panels display a clear positive signal. The VLM outputs a binary prediction (Dwarf/Junk), an associated confidence level, and a free-form explanation.}
\label{fig:figure1}
\end{figure}

The campaign presented 46,334 subjects to more than 1,650 human volunteers. Each subject received multiple (25) independent votes, resulting in over 1.1 million classifications. The subjects include 43,677 candidates from the data, 2,600 simulated galaxies, and 57 known dwarf galaxies. Each subject consists of a set of diagnostic plots (Fig.\ \ref{fig:figure1}), accompanied by instructions to distinguish genuine dwarf galaxy candidates from artifacts (“junk”). Classification requires integrating information across multiple panels, which makes this a more realistic and challenging setting than standard single-image benchmarks.

\paragraph{Models and Prompting:} In this work, we use the multimodal GPT-5-mini model\cite{GPT_5_modelcard} for all experiments. This model provides a practical balance between performance, speed, and cost -- an important consideration in realistic scenarios, where a VLM-based approach must be both accurate and scalable to large candidate sets. We leverage the OpenAI Batch API, which enables asynchronous processing of large classification workloads at reduced cost.

We focus on a zero-shot prompting setup, using text-only instructions that describe how to distinguish genuine dwarf galaxies from artifacts. These instructions closely mirror those provided to human annotators in the citizen science campaign. While VLMs can also incorporate visual examples (few-shot prompting), we defer this direction to future work in order to isolate the capabilities of instruction-only guidance.

All prompts and code used in this work are provided in the accompanying GitHub repository\footnote{\url{https://github.com/dtanoglidis/Zooniverse-Dwarf-Multimodal}}.

\section{Results}

\begin{figure*}[!ht]
\centering
\subfloat[]{\includegraphics[width=59mm]{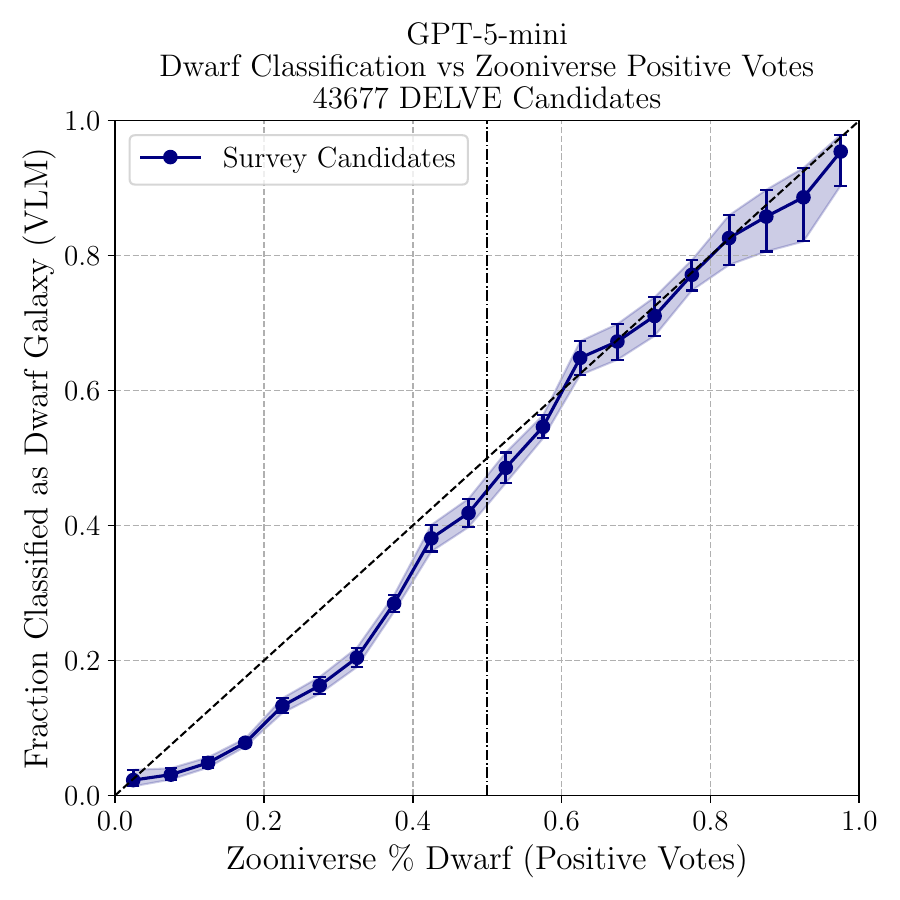}} \quad \quad 
\subfloat[]{\includegraphics[width=57mm]{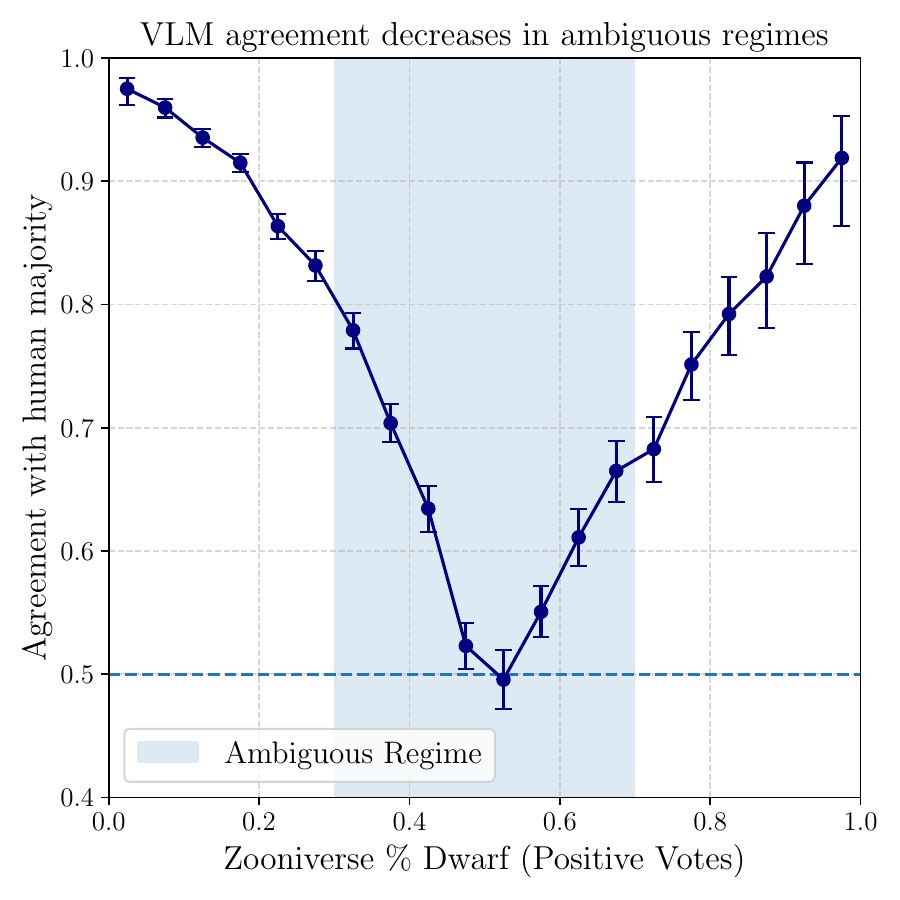}} \quad
\caption{(a) VLM dwarf classification fraction vs. human positive vote fraction, showing strong agreement at the population level. (b) Agreement between VLM and human labels decreases in ambiguous regimes.}
\label{fig:figure2}
\end{figure*}

\begin{figure*}[!ht]
\centering
\subfloat[]{\includegraphics[width=61mm]{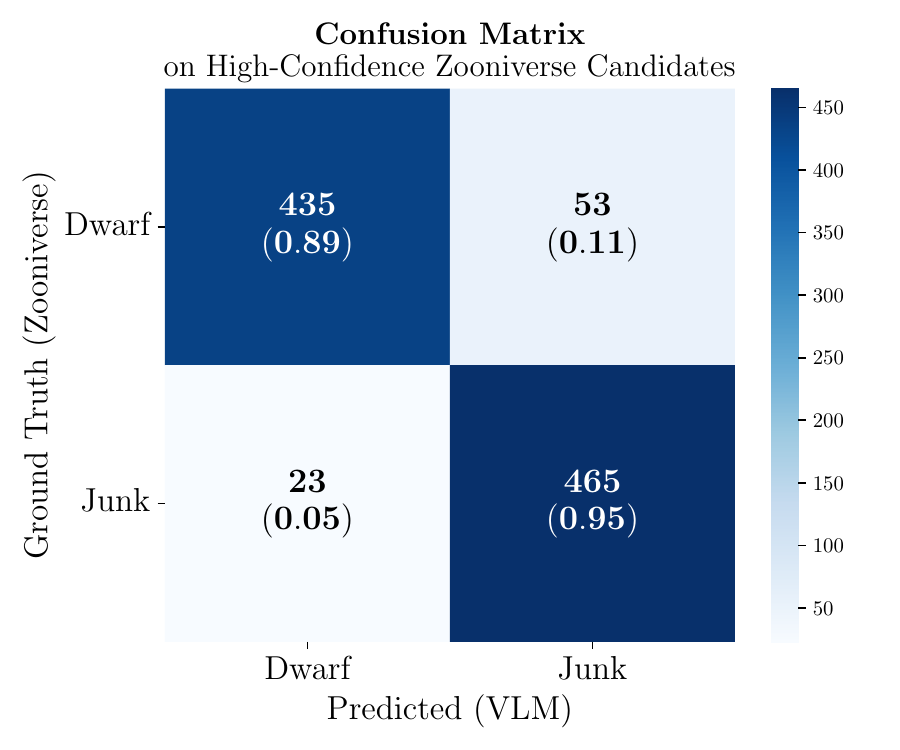}} \quad \quad 
\subfloat[]{\includegraphics[width=49mm]{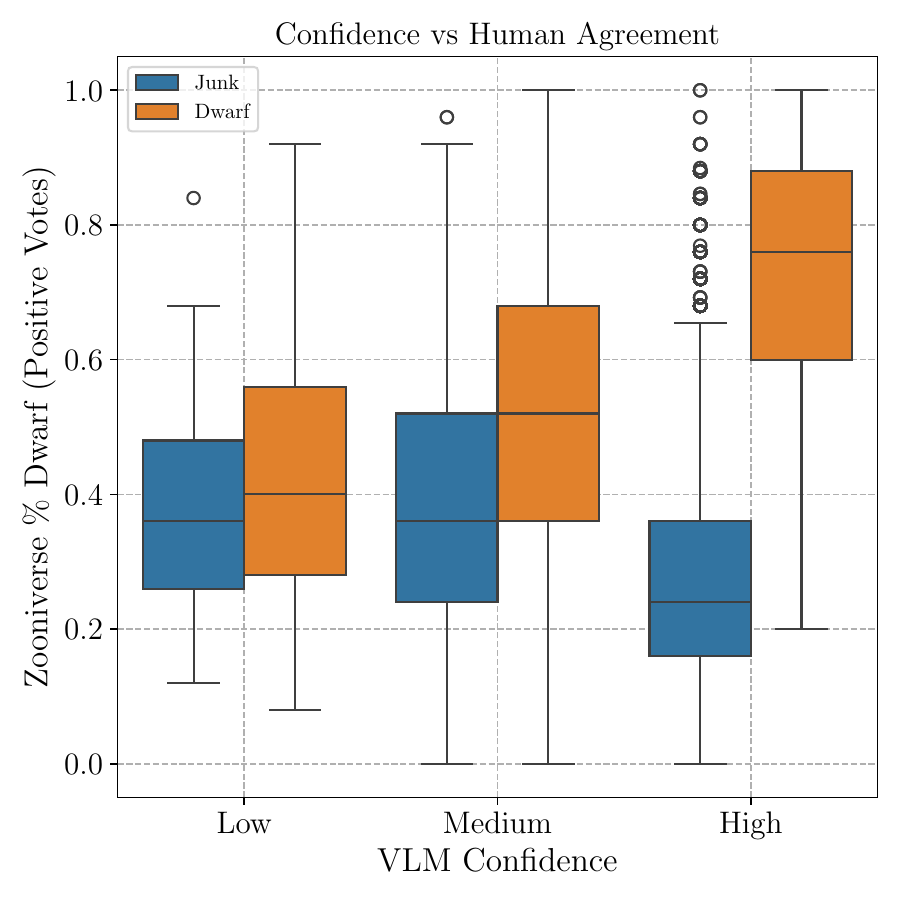}} \quad
\caption{(a) Confusion matrix on high-confidence human-labeled candidates, showing high precision and recall. (b) VLM confidence correlates with human votes, but high-confidence predictions are sparse.}
\label{fig:figure3}
\end{figure*}

Using GPT-5-mini with a zero-shot instruction prompt, we classify the 43,677 DELVE survey candidates, yielding 12,599 labeled as Dwarf. We compare these predictions to the fraction of positive (“is Dwarf”) votes from Zooniverse volunteers (Fig. \ref{fig:figure2}). Binning candidates by human positive fraction (Fig. \ref{fig:figure2}a), the fraction classified as Dwarf by the VLM closely matches the human signal, indicating good aggregate calibration. In Fig. \ref{fig:figure2}b we binarize the human labels (positive fraction $>0.5$ as Dwarf, $<0.5$ as Junk) and compute agreement with the VLM across bins. Agreement is strongest in low-ambiguity regimes (very low/high vote fractions).
Furthermore, the model correctly identifies 55/57 known dwarfs, with the two failures corresponding to challenging cases, even for humans.

To further quantify performance, we construct a high-confidence subset using human votes ($>0.85$ for dwarf, $<0.15$ for junk). On this subset, the VLM achieves $95\%$ precision and $89\%$ recall (Fig. \ref{fig:figure3}a), consistent with prior work on clear, less-ambiguous astronomical classification problems.

\begin{figure*}[!ht]
\centering
\subfloat[]{\includegraphics[width=60mm]{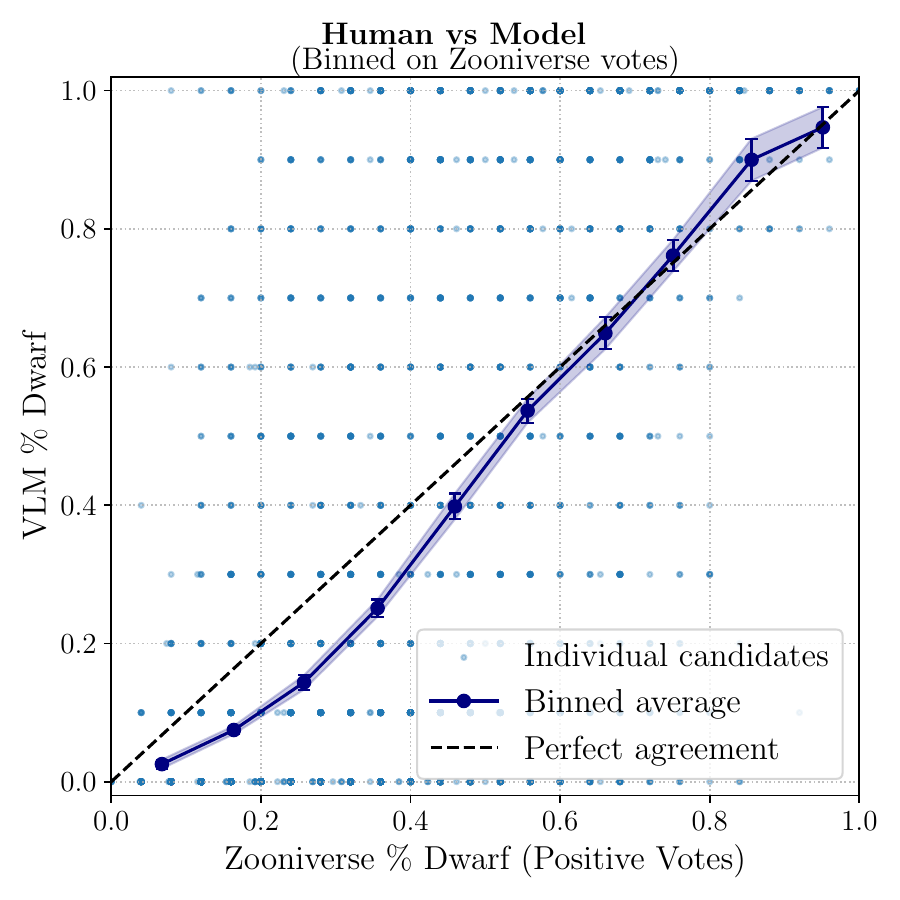}} \quad \quad 
\subfloat[]{\includegraphics[width=60mm]{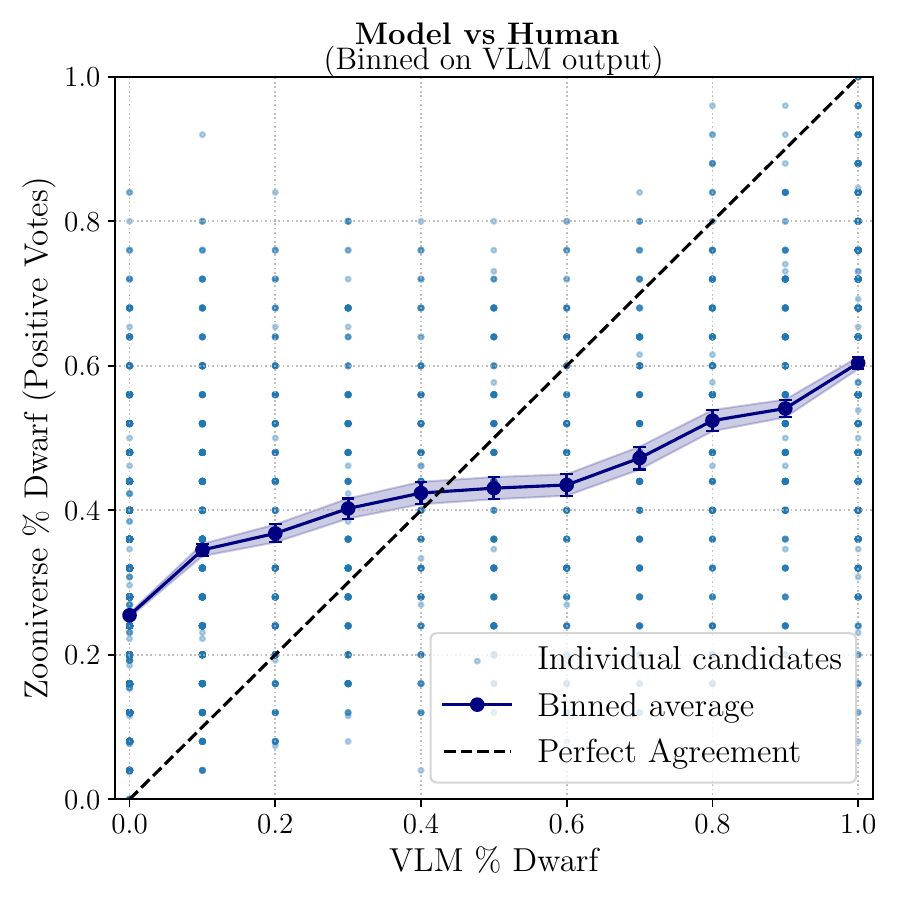}} \quad
\caption{(a) Binned on human votes: model frequency tracks human consensus. (b) Binned on model output: human agreement does not map cleanly to model frequency.}
\label{fig:figure4}
\end{figure*}

For VLM classifiers to be useful in ambiguous and hard cases, it is important to obtain a meaningful and reliable score together with the binary label. Such a score enables targeted follow-up, focusing on the highest-confidence candidates—analogous to selecting objects with high human agreement. In Fig. (\ref{fig:figure3}b), the model’s self-reported confidence correlates with human vote fraction. However, only a small fraction (135 of those classified as Dwarf) of the examples are assigned High confidence, limiting their practical utility.

As an alternative, we estimate uncertainty via repeated inference. We select a random subset of 4,000 candidates (for speed and cost) and run the VLM 10 times per example, defining a model positive fraction (fraction of runs classified as Dwarf).
In Fig. \ref{fig:figure4}, we compare this quantity to human votes. In Fig. \ref{fig:figure4}a, we bin by human positive fraction and compute the mean VLM dwarf frequency in each bin. The solid points show these means, with error bars given by the standard error of the mean ($\sigma/\sqrt{n}$), reflecting the uncertainty on the average within each bin. In this view, the model frequency increases with human consensus, indicating good average calibration.
In contrast, Fig. \ref{fig:figure4}b bins by the VLM output (model positive fraction) and computes the corresponding mean human vote fraction. 
Here, while the mean human vote fraction increases with VLM "Dwarf" classification frequency, the mapping is not well calibrated: each VLM bin contains a broad spread of human votes (low $\%$ and high $\%$), reducing its usefulness as a probabilistic estimate.

Overall, while VLMs reproduce aggregate human trends and perform well on clear cases, their outputs remain poorly calibrated at the level of individual, ambiguous examples.

\section{Discussion and Conclusions}

In this work, we compare the performance of VLMs in discovering ultra-faint dwarf galaxies in survey data against human annotations from a large-scale citizen science campaign. We find that VLMs perform strongly on clear, low-ambiguity cases and reproduce aggregate human behavior, suggesting they can serve as a scalable first-pass filter. However, reliable uncertainty estimation remains a key limitation. Both self-reported confidence and repeated inference fail to provide well-calibrated, useful scores, limiting the direct use of VLM outputs for candidate prioritization and follow-up.

In future work we plan to explore alternative proprietary and open-source VLMs (or even results based on an ensemble of VLM, see for example \citep{VLM_Ensembles_2026}), few-shot \cite{Brown_Few_shot} visual prompting, and improved calibration methods. We also plan to investigate human–AI collaborative workflows in which VLM predictions are used to prioritize candidate review and guide scientific annotation efforts. Overall, while VLMs show clear promise, robust uncertainty and per-example reliability remain open challenges for deployment in scientific workflows.


\medskip

{
\small

\bibliographystyle{plainnat}
\bibliography{ref}

}










\end{document}